\documentclass[aps,prl,twocolumn,groupedaddress,nofootinbib]{revtex4}

\usepackage{graphicx}
\usepackage{amsmath}
\usepackage{bm}

\usepackage{color}

\usepackage{listings}

\begin{document}

\title{A first-principles investigation of the structural and electrochemical properties of biredox ionic species in acetonitrile} 

\author{Kyle G. Reeves, Alessandra Serva, Guillaume Jeanmairet, Mathieu Salanne}
\affiliation{\small Sorbonne Universit\'{e}, CNRS, Physico-chimie des \'Electrolytes et Nanosyst\`emes Interfaciaux, PHENIX, F-75005 Paris \\
\& R\'eseau sur le Stockage Electrochimique de l'Energie (RS2E), FR CNRS 3459, 80039 Amiens Cedex, France}

\begin{abstract}
Biredox ionic liquids are a new class of functionalized electrolytes that may play an important role in future capacitive energy storage devices. By allowing additional storage of electrons inside the liquids, they can improve device performance significantly. However current devices employ nanoporous carbons in which the diffusion of the liquid and the adsorption of the ions could be affected by the occurrence of electron-transfer reactions. It is therefore necessary to understand better the thermodynamics and the kinetics of such reactions in biredox ionic liquids. Here we perform ab initio molecular dynamics simulations of both the oxidized and reduced species of several redox-active ionic molecules (used in biredox ionic liquids) dissolved in acetonitrile solvent and compare them with the bare redox molecules. We show that in all the cases, it is necessary to introduce a two Gaussian state model to calculate the reaction free energies accurately. These reaction free energies are only slightly affected by the presence of the IL group on the molecule. We characterize the structure of the solvation shell around the redox active part of the molecules and show that in the case of TEMPO-based molecules strong reorientation effects occur during the oxidation reaction.
\end{abstract}

\maketitle

\section{Introduction}\label{Intro}

Ionic liquids constitute a versatile family of electrolytes: their structure as well as their physico-chemical properties often vary markedly depending on the ionic species that constitute them~\cite{hayes2015a}. For example, the viscosity can change by almost one order of magnitude by changing the length of the alkali chains in imidazolium-based ionic liquids~\cite{tariq2011a}. As a consequence, there have been high expectations for the use of ionic liquids as electrolytes for energy storage applications~\cite{armand2009a,macfarlane2014a}. Despite an intense activity in the field, their use however remains very limited to niche applications such as the development of multivalent-ion batteries~\cite{koketsu2017a}.

Another interesting characteristic of ionic liquids lies in the fact that they can also be functionalized in order to enhance their efficiency for a specific task such as CO$_2$ capture~\cite{bates2002a}, the extraction of heavy metals~\cite{visser2002a} or redox activity~\cite{chamiot2009a}. Following this idea, Mourad {\it et al.} have recently introduced so-called biredox ionic liquids (BILs), in which both ions are functionalized with a redox-active moiety~\cite{mourad2016a}. An interesting feature of BILs is that the anion can be reduced while the cation can be oxidized, forming multi-charged species rather than neutral ones. This possibly extends their applications into the context of supercapacitors, which are energy storage devices that operate through the adsorption of ions from an electrolyte at the surface of nanoporous carbon electrodes~\cite{simon2008a,salanne2016a}. In conventional systems, the electrons and holes are stored inside the electrode only. When using BILs, additional electrons/holes are accumulated inside the liquid, leading to a significant increase of the capacitance of the system~\cite{mourad2017a,bodin2018a}.

BILs are therefore a promising avenue of research for future supercapacitors. Nevertheless, much remains to be understood about the involved charging mechanisms. In particular, the interplay between ionic diffusion inside the pores, ion adsorption at the surface of the carbon and electron transfer events all remain completely unknown. As established by Fontaine, the first step will consist in acquiring a deeper understanding of the physical chemistry of electron transfers in such media~\cite{fontaine2019a}. Ionic liquids, due to their peculiar structure characterized by Coulomb ordering~\cite{hayes2015a} and eventual formation of polar/nonpolar domains~\cite{canongialopes2006a,kashyap2012a}, cannot be considered as simple dielectric media as is the case of water or simple organic solvents, especially when adsorbed at electrified interfaces~\cite{fedorov2014a}. Nevertheless, several studies which have focused on the impact of this structure on the thermodynamics and kinetics of electron transfers, both by experiments~\cite{tanner2015a,tanner2015b} and simulations~\cite{lyndenbell2007a}, have concluded that the generic concepts introduced by Marcus more than fifty years ago~\cite{marcus1965a}, such as the solvent reorganization energy, may be applicable provided that a few corrections are introduced.

\begin{figure}[t]
\begin{center}
\includegraphics[width=\columnwidth]{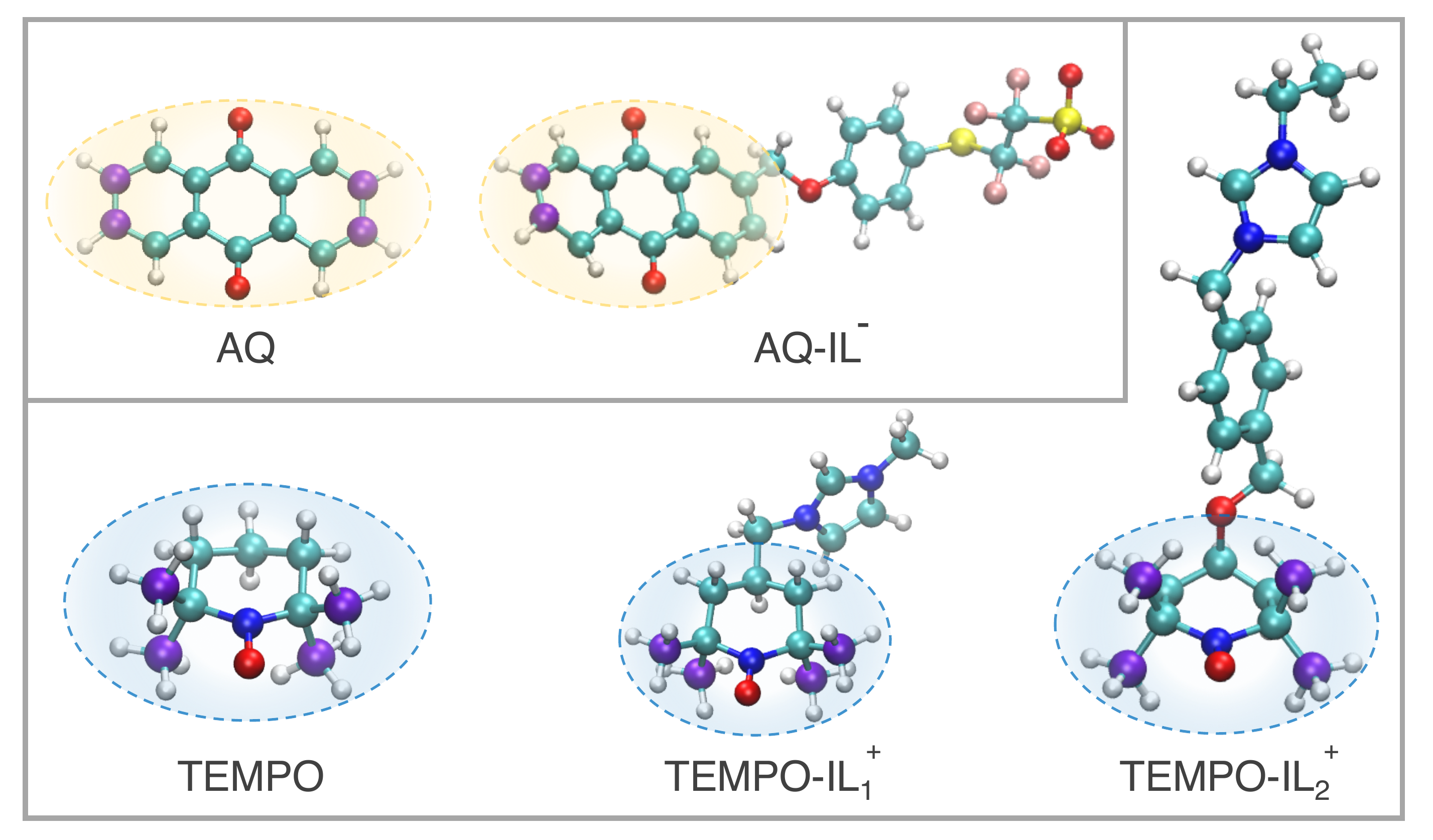}
\end{center}\caption{ The molecules investigated in this work include anthraquinone (AQ), 2-methyloxaphenylperflurosulfonate-anthraquinone (AQ-IL$^{-}$), 2,2,6,6-tetramethylpiperidinyl-1-oxyl (TEMPO), methylimidazolium-TEMPO (TEMPO-IL$_{1}^{+}$), and methylimidazolium-p-xylyloxa-TEMPO (TEMPO-IL$_{2}^{+}$). The redox-active component is highlighted within the coloured region, where yellow represents the part of the molecule to be reduced and blue represents the part of the molecule to be oxidized.}\label{fig:scheme}
\end{figure}

In this work we study independently the redox properties of the ionic species forming BILs by using ab initio molecular dynamics simulations. We first focus on the redox compounds, namely the anthraquinone (AQ) and the 2,2,6,6-tetramethylpiperidinyl-1-oxyl (TEMPO), then we study the functionalized BIL molecules, i.e. the 2-methyloxaphenylperflurosulfonate-anthraquinone (AQ-IL$^{-}$), the methylimidazolium-TEMPO (TEMPO-IL$_{1}^{+}$), and the methylimidazolium-p-xylyloxa-TEMPO (TEMPO-IL$_{2}^{+}$). The structures of all these molecules are provided in Figure \ref{fig:scheme}. In this first step, following the experimental studies of Mourad {\it et al.}~\cite{mourad2016a}, we consider the simplified case where the biredox molecules are dissolved in an organic solvent, acetonitrile. This allows us to put the results obtained in perspective with the experiments, as well as to avoid sampling issues that would arise due to the large viscosity of neat BILs. We follow the approach introduced by Warshel, which consists in using the vertical energy gap between the reactant and the product of a reaction as a reaction coordinate~\cite{warshel1982a}. As was shown by Sprik and co-workers, the vertical energy gap is very appropriate in the context of ab initio molecular dynamics, and it allows to determine accurately redox properties in aqueous and organic solvents~\cite{vuilleumier2001a,blumberger2004a,blumberger2006a,vandevondele2006a,vandevondele2006b}. We show that the ionic liquid functionalization leads to small shifts in both the oxidation potentials and in the reorganization energies. The small variation is easily interpreted by the fact that the structure of the solvent around the redox-active part of the molecules remains generally the same for a given redox moiety.    

\section{Theory and simulations}\label{methods}
\begin{table}[]
\begin{center}
\begin{tabular}{l|lccccc}
\hline 
\vspace{-5pt}\\
Sim &  Solute  & Ion & N   & L (\AA)  & $\langle \Delta E \rangle$ (eV) & $\sigma$ (eV)   \\

\vspace{-5pt}\\
\hline 
\vspace{-0.2pt}\\
1 & TEMPO$^\bullet$ & - & 96 & 20.591& 3.585 & 0.194 \\
\vspace{-9pt}\\
2 & TEMPO$^+$ & - & 96 & 20.591 & 1.478 & 0.235 \\
\vspace{-9pt}\\
\hline
\vspace{-9pt}\\
3 & TEMPO-IL$^{\bullet+}_1$ & Cl$^-$ & 193 & 25.922 & 3.448 & 0.176\\
\vspace{-9pt}\\
4 & TEMPO-IL$^{2+}_1$ & Cl$^-$ & 193 & 25.922 & 1.995 & 0.248\\
\vspace{-9pt}\\
\hline
\vspace{-9pt}\\
5 & TEMPO-IL$^{\bullet+}_2$ & Cl$^-$ & 191 & 25.962 & 3.274 & 0.149\\
\vspace{-9pt}\\
6 & TEMPO-IL$^{2+}_2$ & Cl$^-$ & 191 & 25.962 & 2.045 & 0.239 \\
\vspace{-9pt}\\
\hline
\vspace{-9pt}\\
7 & AQ & - & 95 & 20.550 & -0.246 & 0.194\\
\vspace{-9pt}\\
	8 & AQ$^{-}$ & - & 95 & 20.550 &  0.921 & 0.153\\
\vspace{-9pt}\\
\hline
\vspace{-9pt}\\
9 & AQ-IL$^-$ & Li$^+$ & 189 & 25.970 & -0.281 & 0.211\\
\vspace{-9pt}\\
10 & AQ-IL$^{2-}$ & Li$^+$ & 189 & 25.970 & 0.899 & 0.179\\
\hline
\end{tabular}
\end{center}
	\caption{Summary of the simulation setups and main results. The second column indicates the redox solute which was present in the simulation box, the third one indicates which counter-ion was added, the fourth one is the number of acetonitrile molecules, the fifth one is the length of the cubic box in Angstroms (\AA), the sixth and the seventh column are the mean and the standard deviation of the computed vertical energy gaps.}
\label{tab:sim_details}
\end{table}

 We study five redox half reactions of the form 
\begin{equation}
\text{Ox + e}^-\text{= Red} \label{eq:halfredox}
\end{equation}
\noindent where the two AQ-based molecules displayed in Figure \ref{fig:scheme} are the oxidized form while the three TEMPO-based molecules are the reduced form. Ab initio molecular dynamics simulations were performed for each oxidation state, leading to a total of ten simulated systems for which the compositions and dimensions are displayed in Table~\ref{tab:sim_details} (details of the simulations are provided below). To analyze the electrochemical properties of each system, we computed the vertical energy gap (VEG), defined as the energy difference between the oxidized and reduced species (which are respectively noted with the subscript 1 and 0) for a given solvent configuration, $\textbf{R}^N$, 
\begin{equation}
    \Delta E( \textbf{R}^N ) = E_{1}( \textbf{R}^N ) - E_{0}( \textbf{R}^N ). \label{deltaE}
\end{equation}
This quantity was previously shown to be the proper reaction coordinate to describe electron transfer reactions \cite{warshel1982a}. 
Once the VEGs are properly sampled along the potential energy surface of both the oxidized and the reduced species of a redox couple, it is straightforward to analyze the data in the framework of Marcus theory~\cite{marcus1965a}. The only necessary quantities are the mean and the variance of the VEGs, which are provided for our systems in Table~\ref{tab:sim_details}. However, Marcus theory relies on the assumption that the fluctuations of the solvent are Gaussian and that they are identical for the reactant and the product. As shown in section \ref{Result}, this assumption is not fulfilled in the present systems. We thus resorted to a model that was previously proposed to account for such deviations, the so-called two Gaussian states (TGS) model~\cite{vuilleumier2012a,jeanmairet_chapter_2013}. We recall here those equations that are essential for data analysis and discussion. In the TGS, it is assumed that the solvent degrees of freedom can be partitioned into two different solvation states $S_1$ and $S_2$.  The probability distribution of the VEG associated to each state $S_i$ is assumed to be strictly Gaussian, which leads to a corresponding ``Marcus-like'' expression for the free energy:
\begin{eqnarray}
	W^{S_i}_{\eta}(\Delta E) &=&\frac{(\Delta E -\lambda^{S_i} -\Delta A^{S_i})^2}{4 \lambda^{S_i}} \label{eq:WMeta} \\
	   & & + \frac{k_BT}{2}\log(4\pi k_B T\lambda^{S_i}) \nonumber\\
	   & &+\eta(\Delta E+ \Delta A^{S_i}) \nonumber
\end{eqnarray}
\noindent where $\eta=0$ for the reduced species and $1$ for the oxidized species, $k_B$ is the Boltzmann constant, $T$ is the temperature, $\lambda^{S_i}$ is the reorganization energy and $\Delta A^{S_i}$ is the (full) reaction free energy associated with the solvation state $S_i$.

The Landau free energy of a redox species can then be expressed as a function of the reaction coordinate as 
\begin{equation}
	W_{\eta}(\Delta E)=-k_BT\log \left[ e^{-\beta W^{S_1}_{\eta}(\Delta E)}+ e^{-\beta( W^{S_2}_{\eta}(\Delta E)+A^{S_2})}\right] \label{eq:WTGSeta}
\end{equation}
where $\beta=(k_BT)^{-1}$ and $A^{S_2}$ is a parameter fixing the position of the free energy parabolas associated to state $S_2$ with respect to the ones associated to $S_1$. The TGS therefore introduces five different physical quantities which are treated as parameters, namely $\lambda^{S_1}$, $\lambda^{S_2}$, $\Delta A^{S_1}$, $\Delta A^{S_2}$ and $A^{S_2}$.
The probability distribution of the VEG can then be expressed as
\begin{equation}
p_{\eta}(\Delta E)=e^{\beta A_{\eta}}e^{-\beta W_{\eta}(\Delta E)} \label{eq:peta}
\end{equation}
where we introduced $A_\eta$ the (full) free energy of the reduced and oxidized species:
\begin{equation}
	A_{\eta}=-k_BT\log \left[e^{-\beta \Delta A^{S_1} \eta}+e^{-\beta (  \eta \Delta A^{S_2} +A^{S_2})}\right] \label{eq:A_eta}
\end{equation}

Using equation \ref{eq:A_eta} the total reaction free energy can be computed as $\Delta A=A_1-A_0$.

\subsection{Computational Details}\label{ComputationalDetails}
A total of five biredox systems were simulated in liquid acetonitrile. Classical molecular dynamics simulations were first performed using the DL\_POLY package \cite{dlpoly} to generate the initial atomic configurations to be used to begin the ab initio molecular dynamics simulations. The size of the simulation cell was determined in order to recover the bulk density of acetonitrile. The number of acetonitrile molecules and dimensions of the cubic simulation cell can both be found in Table \ref{tab:sim_details}.

The simulations were performed using the CP2K code via the Quickstep algorithm\cite{hutter2014a, vandevondele2005a}. The Becke exchange and the Lee-Yang-Parr correlation functional\cite{becke1988a, lee1988a} were used, and all atoms of the system were described using Goedecker-Teter-Hutter (GTH) type pseudopotentials\cite{goedecker1996a, hartwigsen1998a}. Kohn-Sham wavefunctions were constructed using a hybrid Gaussian and plane wave scheme, where a basis set of triple-$\zeta$ quality with TZV2P polarization functions was used along with a 280~Ry plane wave density cutoff. To properly describe the radical species in systems 1, 3, 5, 8 and 10, the scaled self-interaction correction was used\cite{perdew1981a,davezac2005a,vandevondele2005b}. Consistently with previous works, we used values of $a=0.2$ and $b=0.0$ for the corresponding parameters~\cite{vandevondele2006a}. The simulations were performed using the Born-Oppenheimer method with a time step of 0.5~fs.  They were performed at a constant volume and temperature of 330~K. Trajectories were gathered for 20~ps, among which the first 2.5~ps were discarded to take into consideration the initial equilibration of the system. Explicit counterions are included only to counterbalance the charge introduced to the system via the ionic liquid moiety. In the case of the change of total charge of the simulation cell due to oxidation/reduction of the biredox molecule, a neutralizing background is introduced through the Ewald summation technique. The VEG was sampled every 25~fs. Its mean $\left \langle \Delta E \right \rangle$ and standard deviation $\sigma$  are reported in Table \ref{tab:sim_details} for all the systems.


\section{Results and Discussions}\label{Result}

\subsection{Redox properties}\label{redox}
\begin{figure}
\centering
\includegraphics[width=\columnwidth]{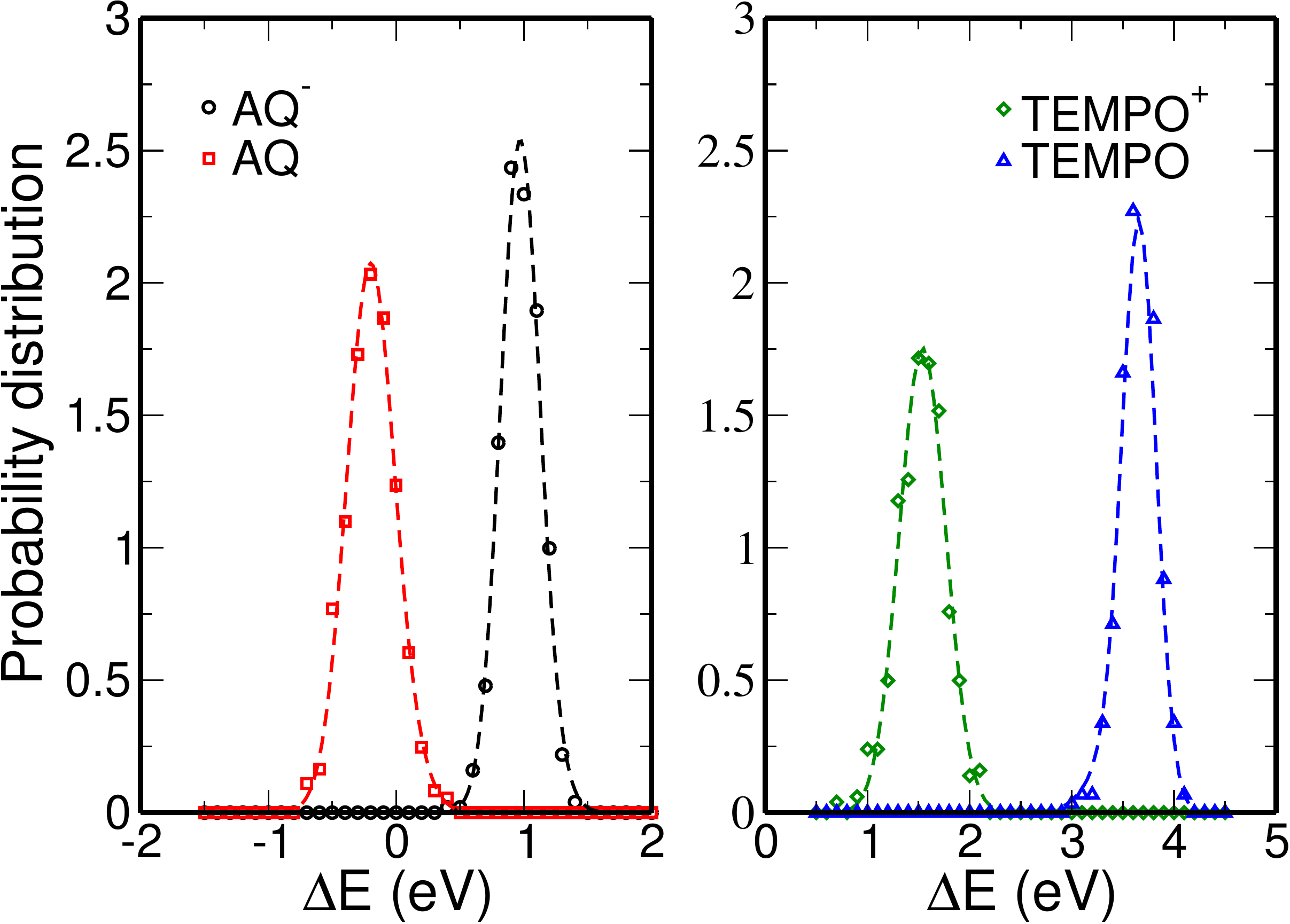}
	\caption{Probability distribution of the energy gap $\Delta E$ of the AQ/AQ$^-$ system (left) and the TEMPO/TEMPO$^+$ system (right). The points computed with ab initio MD are displayed with symbols, the dashed lines were obtained by fitting with the TGS model.}
\label{fig:proba_AQ_TPO_TGS}
\end{figure}

Figure \ref{fig:proba_AQ_TPO_TGS} shows the probability distributions of the VEG for the AQ/AQ$^-$ and the TEMPO/TEMPO$^{+}$ systems.  In both cases, it is quite evident that the data cannot be fitted by a set of two identical Gaussian functions, which indicates that Marcus theory would fail to analyze the results. Deviations are also observed for all the other systems (see Supplementary Figures S1 to S4), and they are especially marked for the TEMPO-functionalized ionic liquid species. We have therefore used the TGS model to analyze all the data. A discussion on the nature of the two solvation states and how they differ between the oxidized and reduced species is provided in subsection \ref{structure}. In practice, the set of associated parameters are fitted on the VEG distributions.

 \begin{figure}
 \protect\centering{}%
 \begin{tabular}{cc}
 \includegraphics[width=\columnwidth]{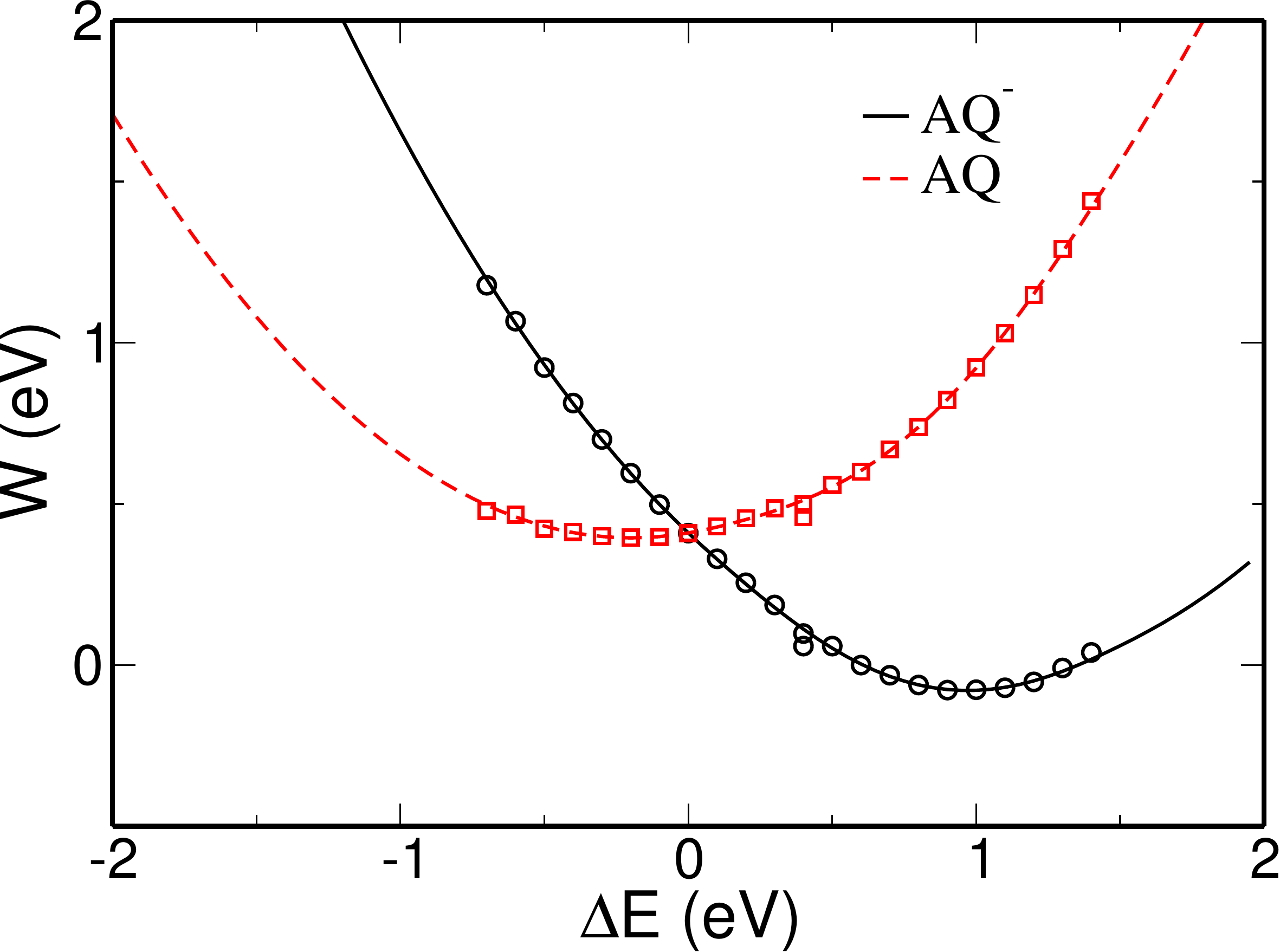} 
 \end{tabular}\protect\caption{Landau free energy curves for the AQ/AQ$^-$ system.
 The raw data from simulations are displayed with symbols, the fitted TGS model using the parameters of Table \ref{tab:TGS_param} are the full and dashed lines. \label{fig:W_AQ_TGS}
}
 \end{figure}

\begin{table}[]
\protect\centering{}%
\begin{tabular}{ c c c c c c }
\hline 
 & AQ & AQ-IL$^-$ & TEMPO  & TEMPO-IL$_1^+$  & TEMPO-IL$_2^+$ \tabularnewline
\hline 
$\lambda^{S_1}$ & 0.616 & 0.665  & 0.910  & 0.844 & 0.674\tabularnewline
$\Delta A^{S_1}$ & 0.416 & 0.376 & 2.449  & 2.888  & 2.710\tabularnewline
$\lambda^{S_2}$ & 0.401 & 0.445  & 0.489  & 0.436 & 0.472 \tabularnewline
$\Delta A^{S_2}$ & 0.564 & 0.401  & 3.172  & 3.057  & 2.851\tabularnewline
$A^{S_2}$ & -0.047 & 0.006 & -0.073 & -0.067  & -0.110 \tabularnewline
\hline 
$\Delta A$ & 0.468  & 0.385  & 2.520 & 2.956  & 2.813 \tabularnewline
\hline 
\end{tabular}\protect\caption{Parameters of the TGS fit (five first lines) and total free energy difference (last line) for the five redox half-reactions (all the values are in eV).
}
\protect
 \label{tab:TGS_param}
\end{table}

 An example of the resulting Landau free energies is given for the system AQ/AQ$^-$ in Figure \ref{fig:W_AQ_TGS} where the lines are  obtained using the fitted TGS model while the symbols correspond to the raw simulation data (obtained with $-k_BT \log p_{\eta}$). A similar good agreement is obtained for all the systems and the set of parameters is given in Table \ref{tab:TGS_param}, together with the resulting half reaction free energies. As shown in previous works~\cite{vandevondele2006a,blumberger2006a}, it is not possible to compare directly these free energies to those determined via experimental because the reference electrode is not the same. Only full reactions, or similar differences in free energies could be compared. In the present case, it is worth noting that experimental information is very scarce, since only half-wave potentials -- which are not true redox potentials -- were reported by Mourad {\it et al.}~\cite{mourad2016a}. Nevertheless these potentials were shown to vary very little with the considered redox species ({\it i.e.} less than 300~mV, which is very small given the large uncertainty in the measure). Our work shows similar results: taking into account an error of the order of 100~meV for each free energy computation (which arises from a statistical uncertainty which was estimated within 50-70~meV by VandeVondele {\it et al.} in similar systems as well as from the typical errors due to the choice of functionals, etc), AQ and AQ-IL can be considered to have the same half reaction free energies. Concerning the series of cationic species, the TEMPO-functionalized ionic liquids show slightly larger free energies than the bare TEMPO, by around 300 to 400~meV. This shows that at least from the thermodynamic point of view, results obtained for the AQ and TEMPO molecules should safely be transferred to the case of BILs.

Concerning the kinetics of the electron transfer reactions, the experimental data (focused on the TEMPO-based molecules) showed that the variations were very small, and that the effective radius that had to be taken into account was the one of the TEMPO moiety (and not the whole molecule) to interpret the data with a modified Marcus theory~\cite{mourad2016a}. In our simulations, we can measure the impact of the solvent on these kinetic aspects by examining the variation of the reorganization energies. We have two sets of data due to the use of the TGS model to build the free energy curves, but we immediately see that the variations are most important for $\lambda^{S_1}$. Indeed, this quantity varies from 0.910~eV for TEMPO to 0.674~eV for the TEMPO-IL$_{2}$ system. This variation will be discussed in terms of solvent molecules residence time in the next section. When comparing the AQ-based and the TEMPO-based system, we also observe a large difference in this reorganization energy, which may hint towards a different variation of the solvation shell during the redox reaction. This point will also be discussed below.

\subsection{Localization of the additional electron/hole}\label{electronicstructure}

\begin{figure}[]
\begin{center}
\includegraphics[width=\columnwidth]{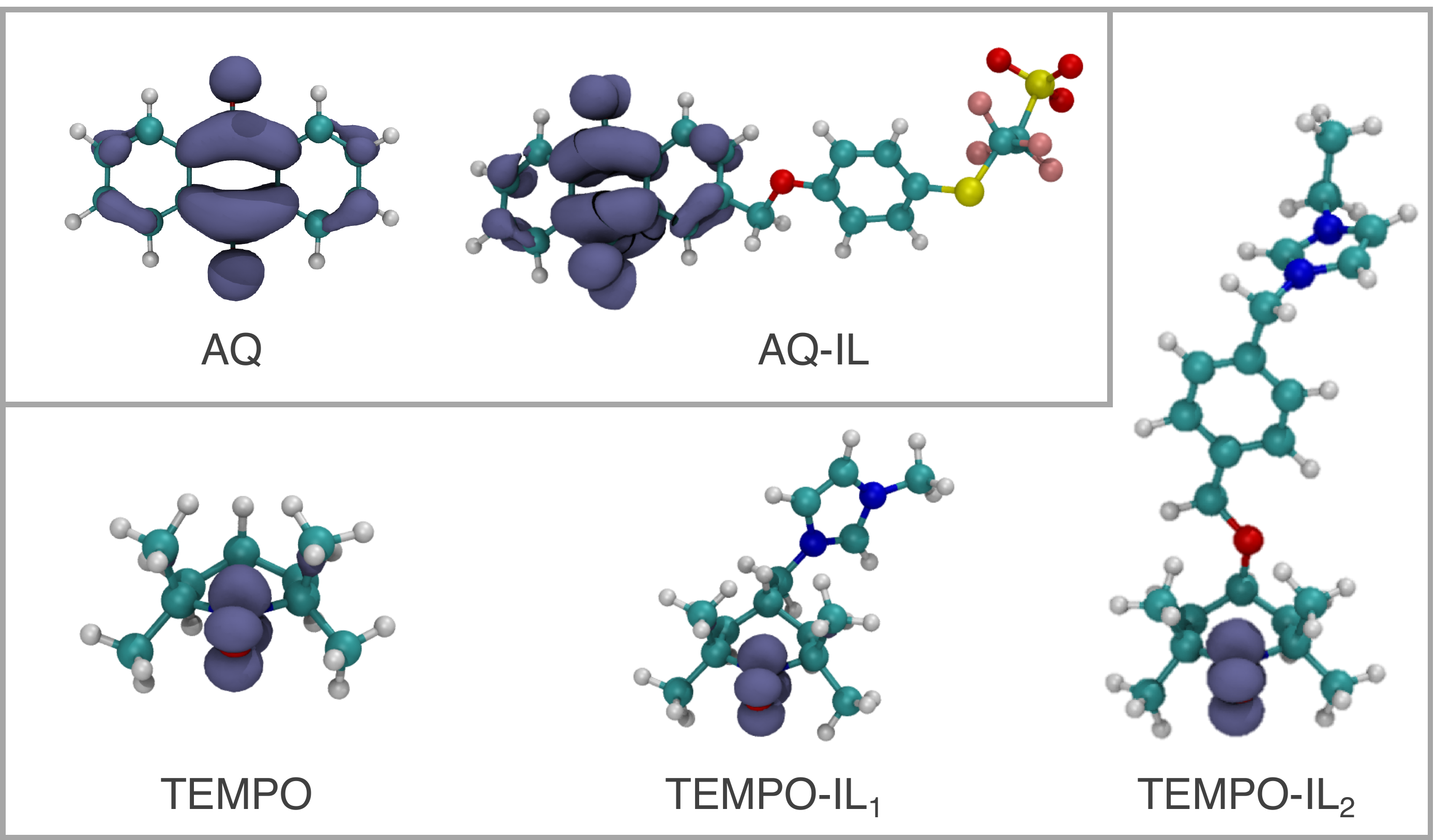}
\end{center}\caption{Electron/hole localization from DFT for biredox molecules. Top row shows the electron density (gray) for AQ$^{-}$ and AQ-IL$^{2-}$ and the bottom row shows the hole density in TEMPO$^{+}$, TEMPO-IL$_{1}^{2+}$ and TEMPO-IL$_{2}^{2+}$.}\label{fig:dft_1}
\end{figure}

Before focusing on the solvation properties, we first look at the localization of the additional electron following the reduction of AQ to AQ$^-$, shown in Figure \ref{fig:dft_1}. We can see that it neither falls on a single atom of the redox species nor is delocalized evenly over the entire molecule, but it is rather delocalized over the two carbonyl groups for both AQ$^{-}$ and AQ-IL$^{2-}$ with additional electron density on the alpha carbons. It should be emphasized that the IL does not participate nor significantly influence the localization of the excess electron. 

A slightly different picture is observed in the case of TEMPO and the TEMPO-ILs.
Figure \ref{fig:dft_1} shows that the hole density for TEMPO$^{+}$ following oxidation is very strongly localized, on the N-O bond. This is contrary to the case of AQ where there was a delocalisation of the charge on various bonds, notably the two carbonyl groups. A similar localization on the N-O bond is observed for both TEMPO-IL$_{1}^{2+}$ and TEMPO-IL$_{2}^{2+}$. These results will allow us to perform a targeted structural analysis on the solvation shell, by focusing on the regions where the electron density is most affected by the electron transfer reaction.

\subsection{Structural properties}\label{structure}
\begin{figure}[t]
\begin{center}
\includegraphics[width=\columnwidth]{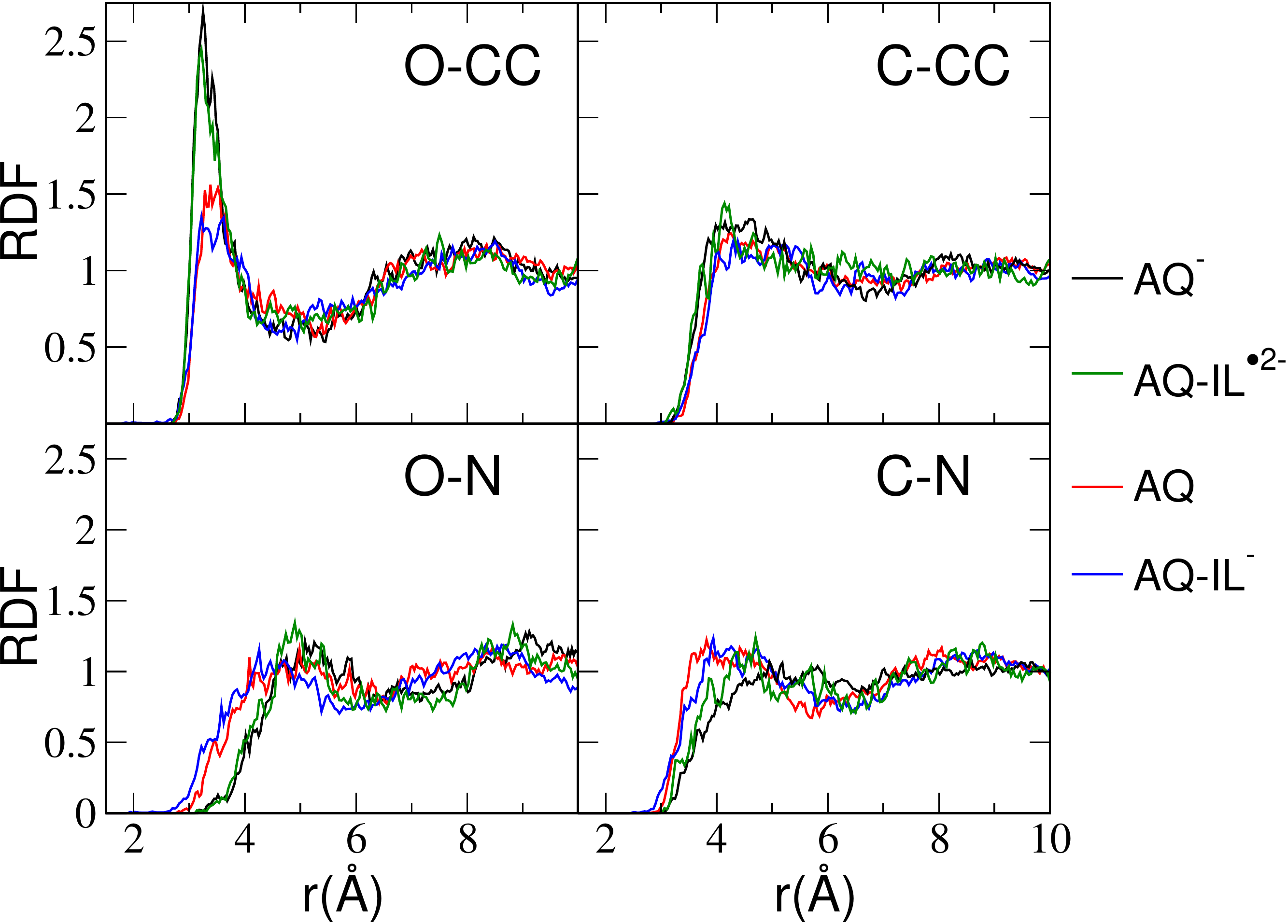}
\end{center}\caption{Selected solute-solvent radial distribution functions (RDFs) calculated for AQ and AQ-IL, for both oxidized and reduced species. The RDFs around the carbonyl oxygen atoms (O) of the AQ moiety are shown in the left panels, while the right panels show the RDFs around selected AQ carbon atoms (C), coloured in violet in Figure \ref{fig:scheme}. For the acetonitrile the CC (methyl carbon) and N atoms have been considered.}\label{fig:rdf_1}
\end{figure}

To understand the molecular origin of the redox properties reported in the previous section, we study the solvation structure of the AQ and TEMPO-based species. First, we report several radial distribution functions (RDFs) between acetonitrile and selected sites of the AQ and AQ-IL systems in Figure \ref{fig:rdf_1}. On the acetonitrile side, we focus on the methyl carbon (CC) and nitrogen (N), while the carbonyl oxygen and the carbon atoms on the outer rings, furthest from the carbonyl group (identified as violet atoms in Figure \ref{fig:scheme}) were selected for the AQ moiety.  We note the general trend that the first peaks in the RDFs appear at long distances (more than 3~\AA) and are not very intense. We can therefore conclude that before reduction, neither of the redox-active species is strongly solvated by the acetonitrile. A similar result has been obtained in a previous study of redox tetrathiafulvalene and thianthrene species in the same solvent~\cite{vandevondele2006a}.

The two different sites of the AQ moiety were chosen in order to observe the impact of electron transfer on the solvent structure both close and relatively far to the region where the electron density is accumulated upon reduction. We immediately see that the interaction between the acetonitrile and the carbon atoms is barely affected, which confirms that most of the solvent reorganization occurs in the vicinity of the carbonyl groups (note that similar conclusion can be drawn from the RDFs involving atoms from the ionic liquid moiety in AQ-IL, that do not change as shown in Figure S5 of the Supplementary Information). Moving to the oxygen atom, the initial organization of the acetonitrile around AQ and AQ-IL$^-$ species, given by the first peak of the RDFs, is characterized by shorter O-CC distances (compared to O-N) and a larger intensity. This is not surprising since acetonitrile is a highly polar molecule (the dipole moment in the bulk is close to 5~D) so that its positive end is attracted by the negatively charged oxygen atoms. When AQ (AQ-IL$^-$) is reduced in AQ$^{-}$ (AQ-IL$^{2-}$), the addition of the electron along the carbonyl bonds reinforces this orientation, leading to an increase of the intensity of the first peak of the O-CC RDF, and consequently to slightly longer O-N distances. No noticeable difference is observed between the bare AQ and the functionalized ionic liquid, which is coherent with the very similar sets of parameters which were obtained when studying the properties of the corresponding half reactions.  

\begin{figure*}[t]
\begin{center}
\includegraphics[width=0.8\textwidth]{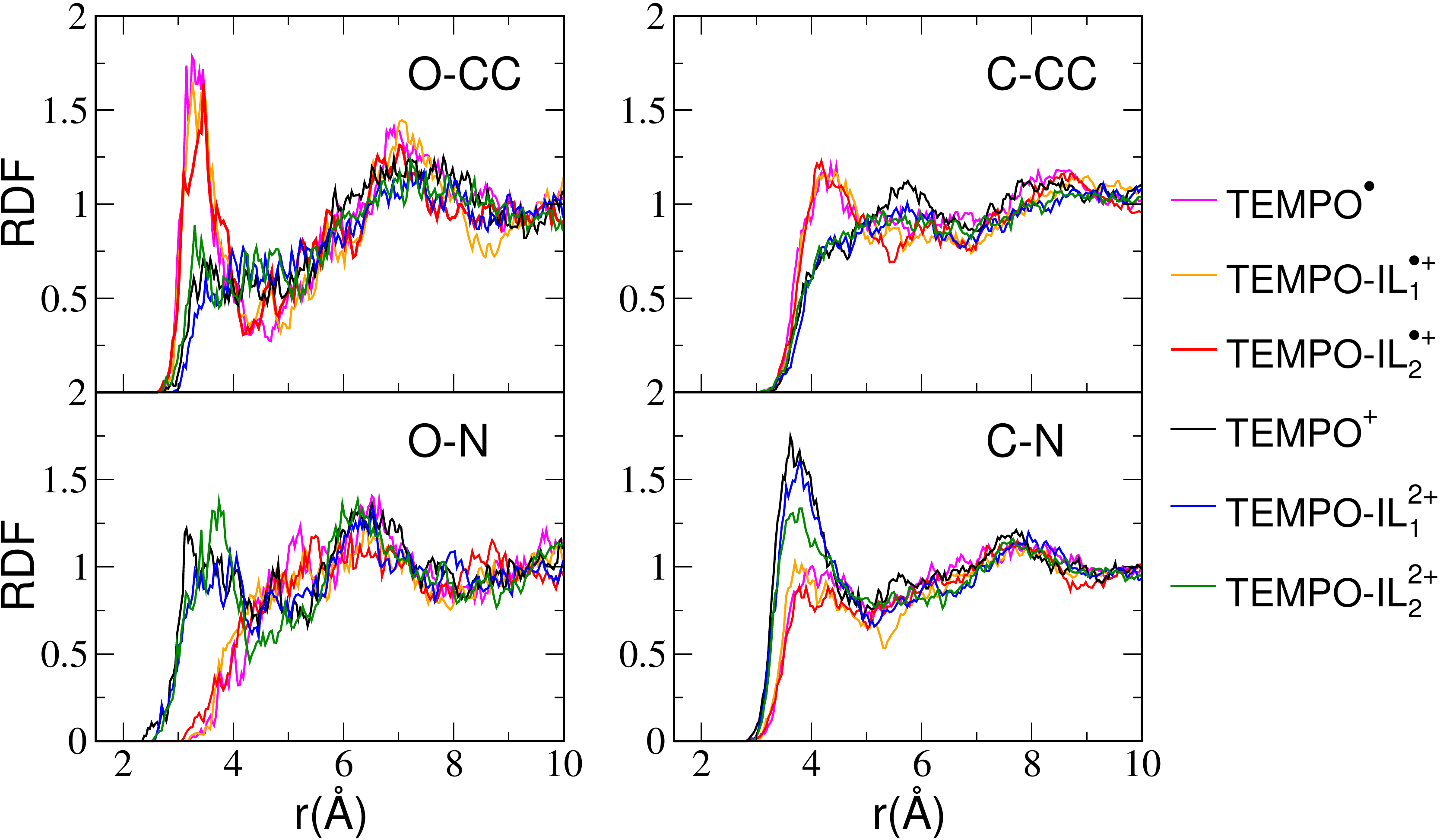}
\end{center}\caption{Solute-Solvent radial distribution functions (RDFs) calculated for TEMPO and TEMPO-ILs, for both oxidized and reduced state. The RDFs around the oxygen atoms (O) of TEMPO moiety are shown in the left panels, while the right panels show the RDFs around selected TEMPO carbon atoms (C), coloured in violet in Figure \ref{fig:scheme}. For the acetonitrile the CC (methyl carbon) and N atoms have been considered.}\label{fig:rdf_2}
\end{figure*}

Similar RDFs are plotted for TEMPO/TEMPO-ILs in Figure \ref{fig:rdf_2}. We selected the oxygen (N-O) and carbon atoms of the four methyl group adjacent to the N-O bond (identified in Figure \ref{fig:scheme}) of the TEMPO moiety, again with the CC and N sites of the acetonitrile. As for AQ-based systems, the initial (here the reduced) species interact very weakly with the acetonitrile molecules, as can be seen from the small intensity of the RDFs first peak. As for the AQ, the solvent molecules are oriented with their methyl group located closer to the oxygen atom of the TEMPO moiety. Marked differences are observed upon oxidation and formation of a positively charged TEMPO group.  The preferential orientation of the acetonitrile methyl group towards the N-O bond (with the nitrogen pointing away from it) for the reduced species is replaced by a mixed solvation shell, with some acetonitrile molecules now oriented with the nitrogen atom towards the N-O bond and with some remaining solvent atoms oriented with the methyl carbon nearest to the bond. In addition, when comparing AQ and TEMPO, the solvent molecules seem to interact more strongly with the methyl groups of the TEMPO as can be seen from the increase of the intensity of the first peaks of the corresponding RDFs.

\begin{table}
\begin{center}
\begin{tabular}{lccccc}
\hline 
\vspace{-7pt}\\
& N$_\text{O-CC}$  & cutoff (\AA) & N$_\text{O-N}$ & cutoff (\AA) & N$_\text{tot}$\\
\vspace{-7pt}\\
\hline 
\vspace{-5pt}\\
TEMPO$^\bullet$ & 2.27 & 4.30 & - & - & 2.27\\
\vspace{-9pt}\\
TEMPO-IL$^{\bullet+}_1$ & 2.26 & 4.30 & - & - & 2.26\\
\vspace{-9pt}\\
TEMPO-IL$^{\bullet+}_2$ & 2.20 & 4.30  & - & - & 2.20 \\
\vspace{-5pt}\\
\hline
\vspace{-7pt}\\\
& N$_\text{O-CC}$  & cutoff (\AA) & N$_\text{O-N}$ & cutoff (\AA) & N$_\text{tot}$\\
\vspace{-7pt}\\
\hline 
\vspace{-5pt}\\
TEMPO$^+$ & 0.84 & 3.90 & 2.65 & 4.40 &3.49 \\
\vspace{-9pt}\\
TEMPO-IL$^{2+}_1$ & 0.64 & 3.90 & 2.40 & 4.40 & 3.04 \\
\vspace{-9pt}\\
TEMPO-IL$^{2+}_2$ & 0.89 & 3.90 & 2.50 & 4.40 & 3.39 \\
\vspace{-5pt}\\
\hline 
\end{tabular}
\end{center}
\caption{Average coordination numbers of the acetonitrile around the oxygen atoms calculated for TEMPO and TEMPO-ILs, for both the reduced (top) and oxidized (bottom) states. CC and N are the methyl carbon and nitrogen acetonitrile atoms, respectively. Note that the cutoff distances have been chosen as the position of the RDFs first minimum.}
\label{ncoord}
\end{table}

Although the orientation of the solvent molecules is likely to influence the reorganisation energies, previous works have shown that the main structural signature of deviations from Marcus theory generally are variations in the solvation numbers~\cite{vuilleumier2012a,hartnig2001a,li2017j,jeanmairet2019a}. Integrating the number of solvent molecules in the first solvation shell, we report in Table \ref{ncoord} the number of acetonitrile molecules around the N-O bond of the redox-active molecule. It increases markedly between the reduced and the oxidized state. There are roughy two acetonitrile molecules (with the methyl group coordinating, as expected from the analysis of the RDFs) for the reduced state and more than three molecules for the oxidized species. In the latter case, on average more than two molecules are oriented with the nitrogen atom towards the N-O bond, and less than one molecule oriented with the methyl group. Again, this behaviour was observed also to be true for the two TEMPO-ILs systems. We can therefore conclude that the strong deviation from Marcus theory observed in this series of systems is linked to this strong change in the solvation shell.

\begin{figure}[t!]
\begin{center}
\includegraphics[width=\columnwidth]{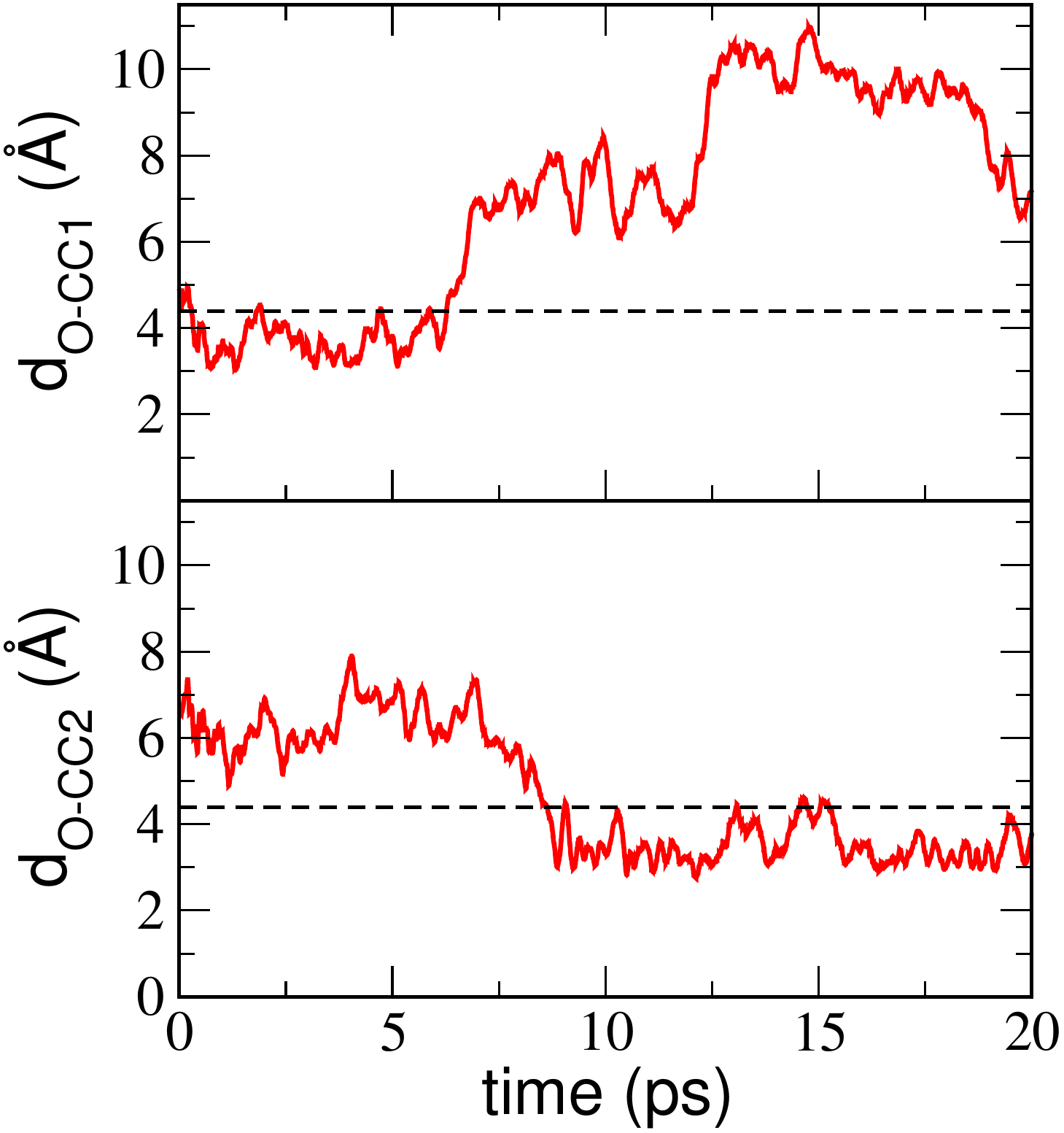}
\end{center}\caption{Time evolution of  the distance between two different acetonitrile molecules (more precisely their two  methyl group carbons, labeled CC1 and CC2) and the oxygen atom O from the TEMPO molecule.}\label{timedistance}
\end{figure}

 If we go back to section \ref{redox}, another interesting result was also the progressive decrease in the reorganization energy $\lambda^{S_1}$ when going from TEMPO to TEMPO-IL$_1$ and finally to TEMPO-IL$_2$, all in the reduced state. Although at this stage we cannot easily comment on the impact of this effect on the kinetics of the redox reaction, we can observe whether these systems differ from the dynamical point of view. Indeed, we observed several exchanges of solvent molecules between the first and the second solvation shells; an example of a molecule arriving and another one leaving is shown in Figure \ref{timedistance} for TEMPO. We can see that different solvation configurations are sampled during the simulation, with the acetonitrile molecules spending several picoseconds in the first shell before leaving it. In particular, the average residence time of the solvent molecules in the TEMPO oxygen first solvation shell has been evaluated by means of the Impey method,~\cite{impey1983a} with  $t^*=100$~fs. Note that $t^*$ is a characteristic time introduced to take account of molecules which leave the first coordination shell only temporarily and return to it. The results show that there is a decrease in the residence time in the direction TEMPO (3.97 ps), TEMPO-IL$_1$ (2.72 ps) and finally TEMPO-IL$_2$ (2.50 ps), i.e the same direction as for the decrease in the reorganization energy. This suggests a less rigid and more dynamical first solvation shell for the functionalized species, thus resulting in a smaller reorganization energy during the electron transfer process. %


\section{Conclusion}\label{conclusion}

This ab initio molecular dynamics study provides a first step towards the understanding of biredox ionic liquids as novel electrolytes for supercapacitors with enhanced performances. By studying a series of TEMPO and anthraquinone-functionalized ionic species dissolved in acetonitrile, we have shown that their electrochemical properties are very similar to the parent redox groups as bare molecules. All the systems cannot be well analyzed by using the standard Marcus theory due to the different solvation shells between the reduced and oxidized species, and thus we introduced a two-Gaussian state model to explore the free energies of the redox reactions. The similar redox properties are easily understood by analyzing the structure of the solvation shell around the functional group: it is barely affected by the presence of an ionic moiety, so that most of the 
difference will probably arise from interfacial effects. The latter will be studied in future works, that will need to involve well-parameterized classical molecular dynamics study for introducing an explicit electrode in the system. Several recent studies have shown that the mechanisms of electron transfer can be affected by such an interface since it may modify again the solvation shell~\cite{remsing2015a,li2017j}. Future work will also be directed towards the analysis of the transport properties of biredox ionic liquids, since knowing quantities such as the diffusion coefficients of the species adsorbed inside electrified nanopores is necessary to assess the power performance of the corresponding supercapacitors~\cite{pean2015b}.

\section*{Conflicts of interest}
There are no conflicts to declare.

\section*{Acknowledgements}
We would like to dedicate this article to Michiel Sprik, who pioneered the study of electrochemical reactions using ab initio molecular dynamics simulations, on the occasion of his 67$^{\rm th}$ birthday and retirement. This project has received funding from the European Research Council
(ERC) under the European Union's Horizon 2020 research and
innovation programme (grant agreement No. 771294). This work was supported by the French National Research Agency (Labex STORE-EX, Grant No. ANR-10-LABX-0076), and it was granted access to the HPC resources of CINES under the
allocations 2018--A0040910463 and 2019--A0060910463 made by GENCI.




\bibliography{references} 

\section*{Supplementary Information}

\subsection*{Probabilities and Landau Free energies}
We report here the probability distribution (top) and the Landau free energies (bottom). The symbols correspond to the data computed by molecular dynamics simulations while the dashed lines correspond to the TGS model with the parameters given in Table 2 of the main article.

 \begin{figure}[h!]
\begin{center}
 \begin{tabular}{cc}
 \includegraphics[width=0.9\columnwidth]{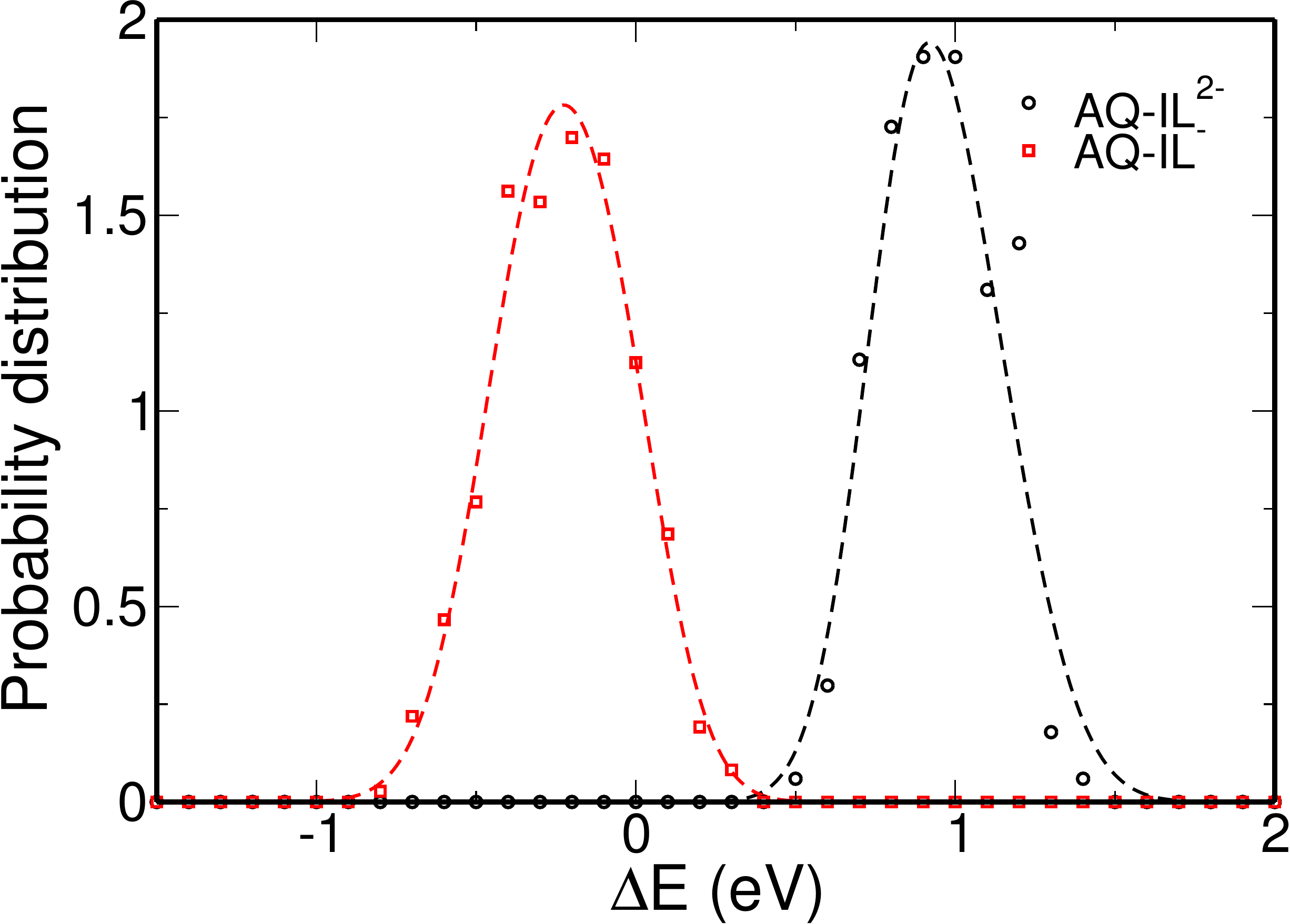}\\ 
 \includegraphics[width=0.9\columnwidth]{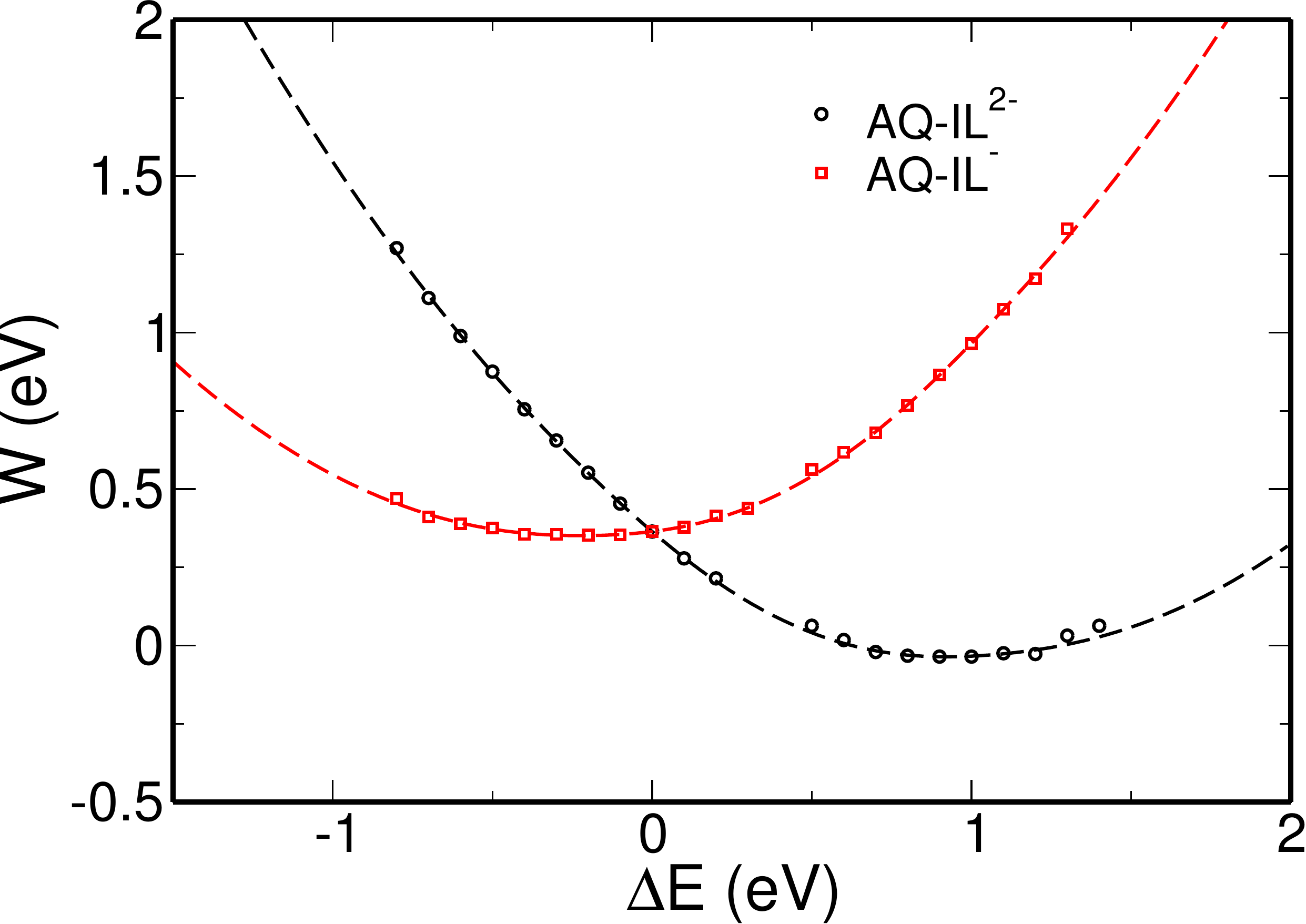} 
 \end{tabular}
\end{center}
	 \figurename{\ S1 Probability distributions and Landau free energy curves for  AQ-IL$^-$/AQ-IL$^{2-}$.
 \label{fig:W_AQ_TGS}
}
 \end{figure}

\begin{figure}[h!]
\begin{center}
\includegraphics[width=0.9\columnwidth]{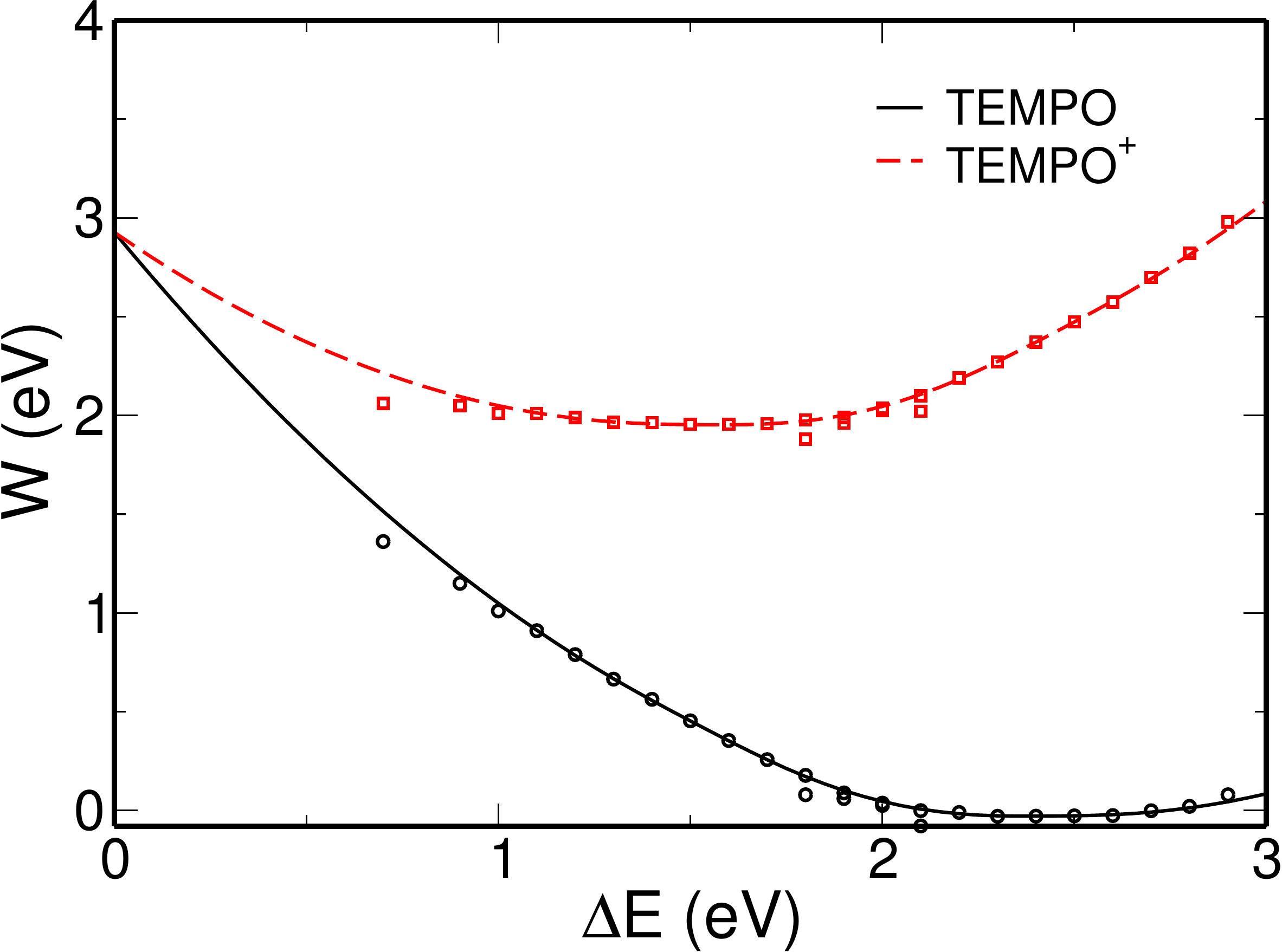}
\end{center}\figurename{\ S2 Landau free energy curves of TEMPO and TEMPO$^+$. The corresponding probability distribution is  displayed in Figure 2 of the main article.}
 \label{fig:W_AQ_TGS}
 \end{figure}

 \begin{figure}[h!]
 \begin{center}%
 \begin{tabular}{cc}
 \includegraphics[width=0.9\columnwidth]{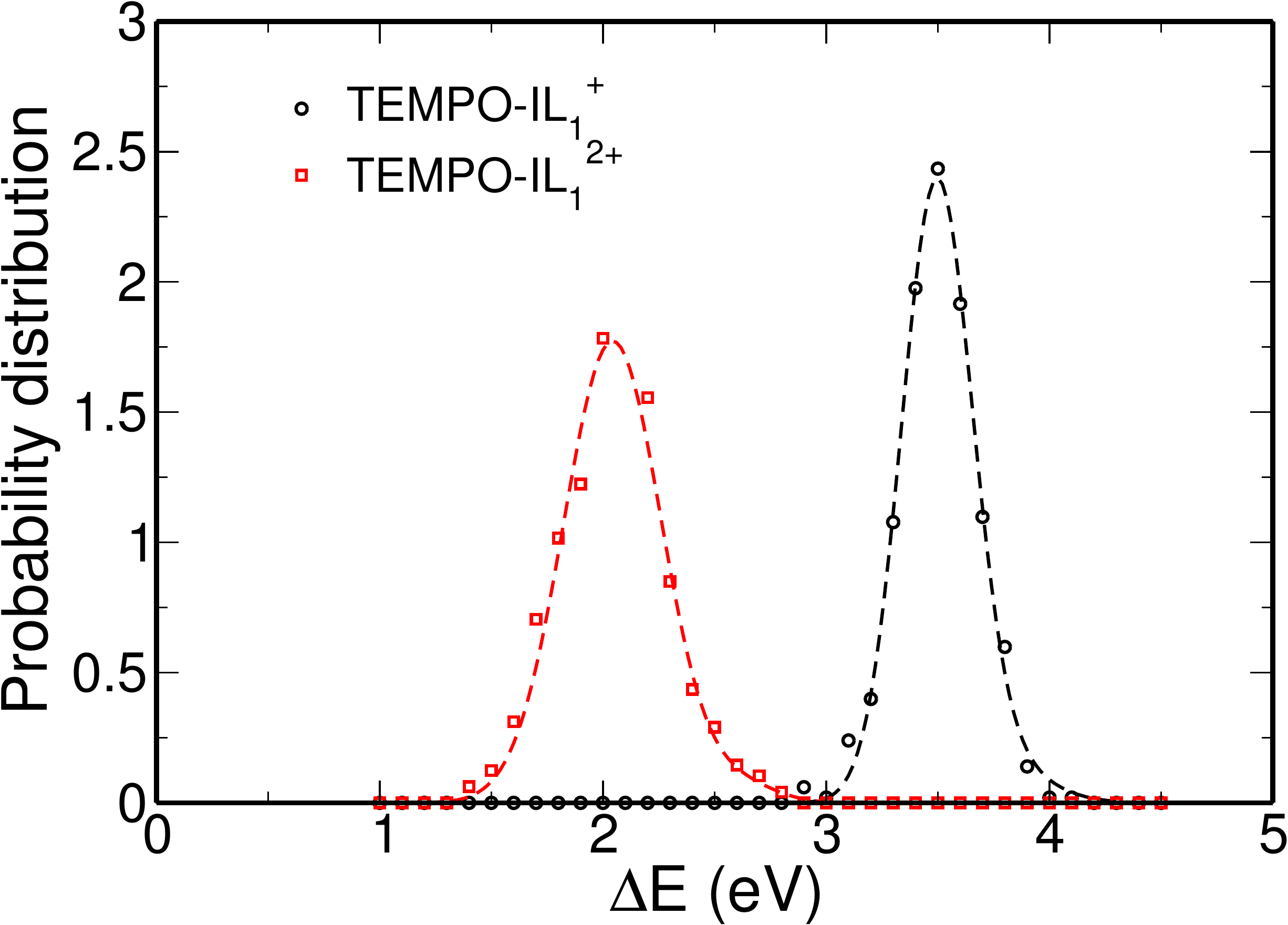}\\ 
 \includegraphics[width=0.9\columnwidth]{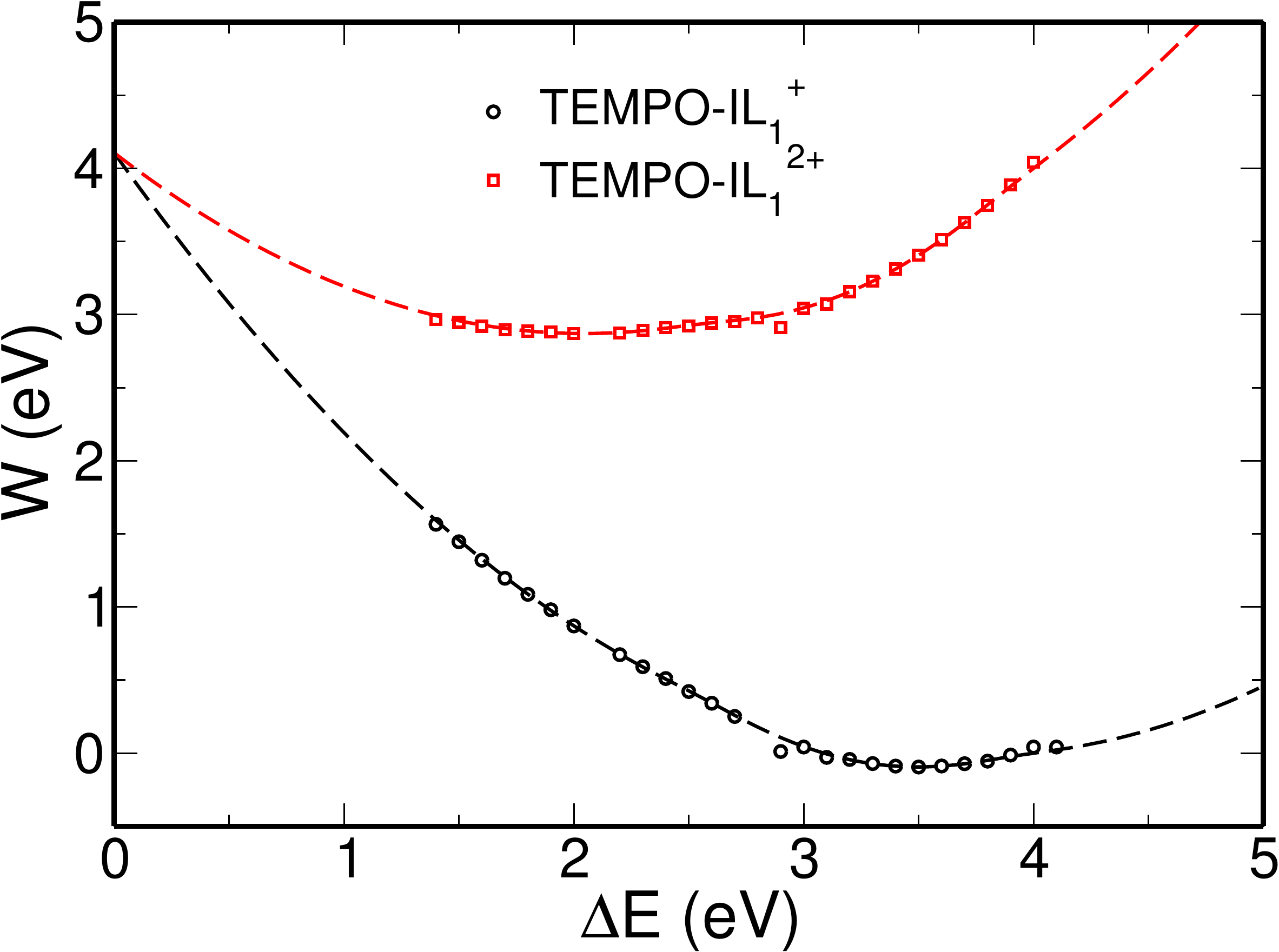} 
 \end{tabular}
 \end{center}	 \figurename{\ S3 Probability distributions and Landau free energy curves for  TEMPO-IL$_1^{\dot +}$/TEMPO-IL$_1^{2+}$.
 \label{fig:W_AQ_TGS}
}
 \end{figure}

 \begin{figure}[h!]
 \begin{center}%
 \begin{tabular}{cc}
 \includegraphics[width=0.9\columnwidth]{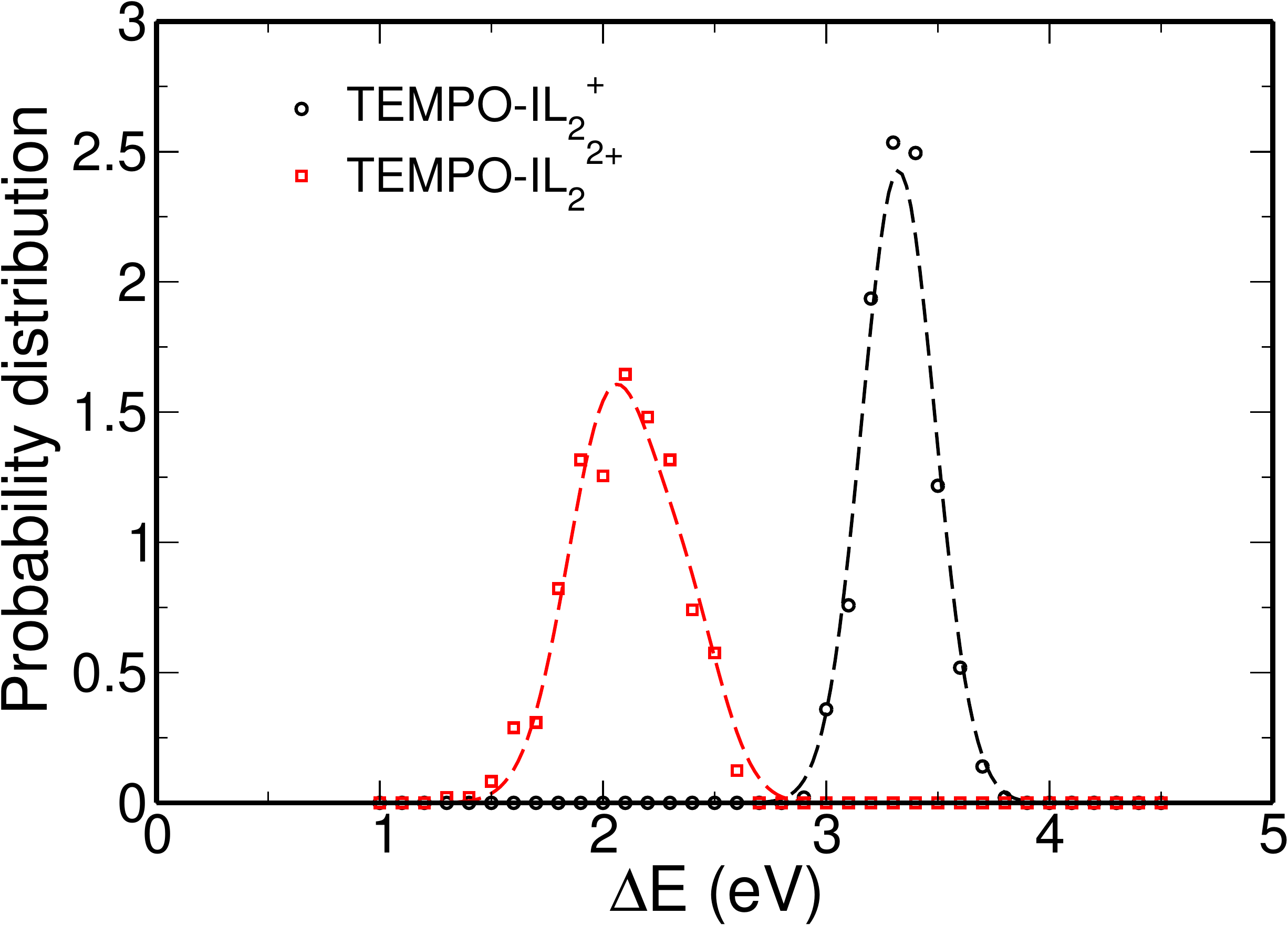}\\ 
 \includegraphics[width=0.9\columnwidth]{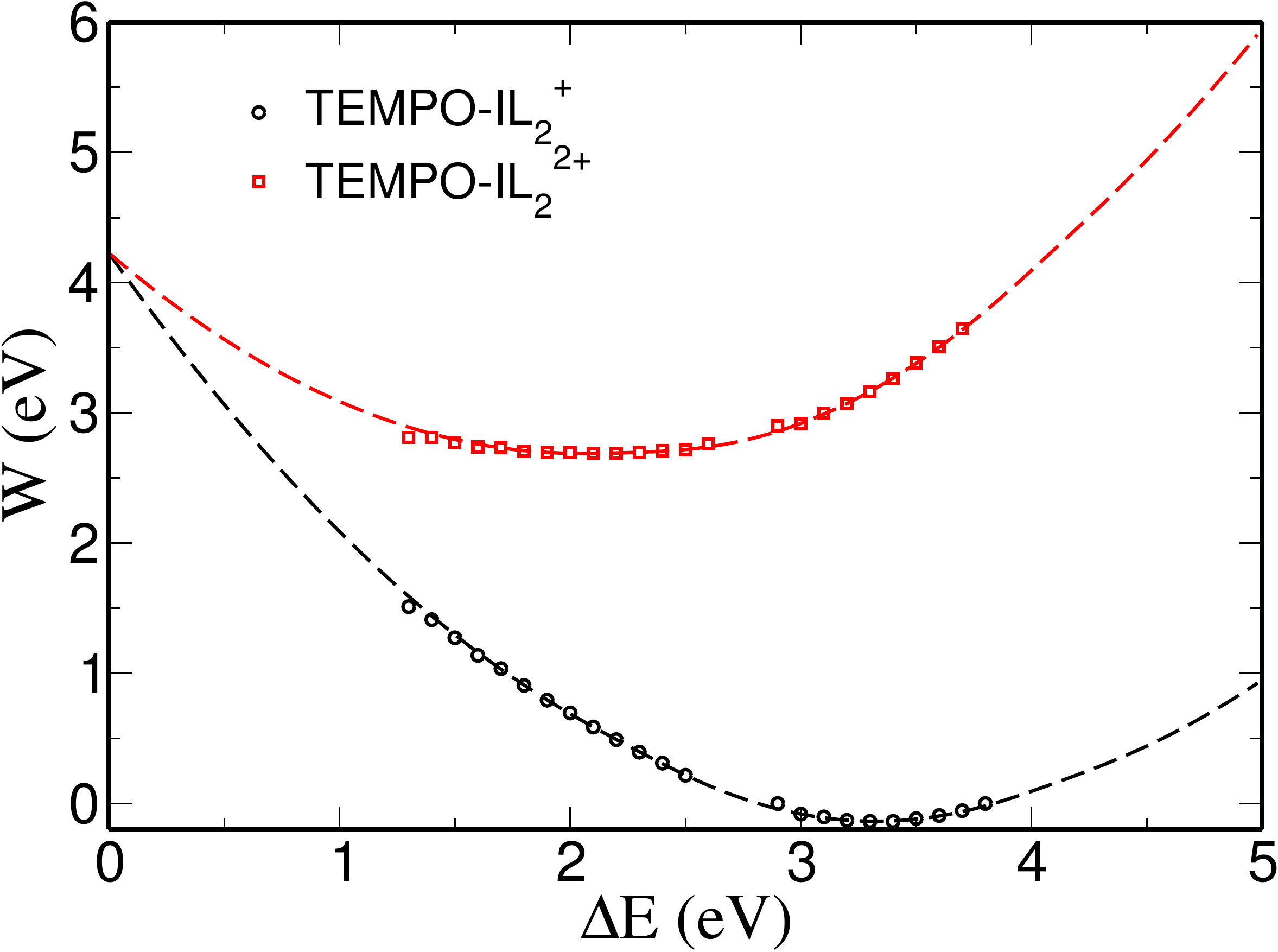} 
 \end{tabular}
 \end{center}
	 \figurename{\ S4 Probability distributions and Landau free energy curves for  TEMPO-IL$_2^{\dot +}$/TEMPO-IL$_2^{2+}$.
 \label{fig:W_AQ_TGS}
}
 \end{figure}

\subsection*{Radial distribution functions}
We report here the RDFs between selected acetonitrile atoms and the oxygen atoms from the ionic liquid moiety, extracted from the AQ-IL$^-$ and AQ-IL$^{\cdot 2-}$ simulations.

\begin{figure}[h!]
\begin{center}
\includegraphics[width=0.9\columnwidth]{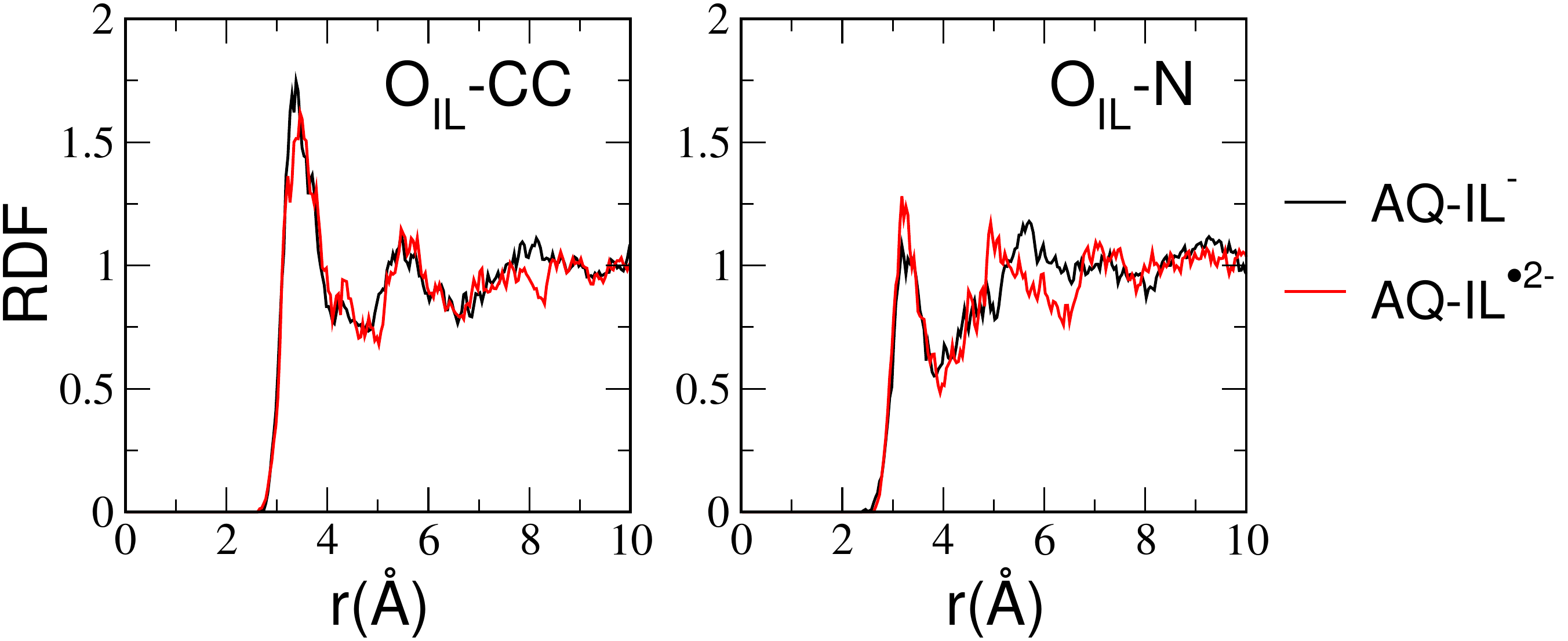}
\end{center}
\figurename{\ S5. O$_{IL}$-CC (left) and O$_{IL}$-N radial distribution functions (RDFs). CC and N are the methyl carbon and nitrogen atoms of the acetonitrile, respectively. O$_{IL}$ are the three oxygen atoms of the SO$^{-}_{3}$ group belonging to the IL part.}
\label{fig:}
\end{figure}

\end{document}